\documentclass[preprint,12pt]{aastex}
\usepackage{natbib}
\usepackage{epsfig}
\begin{document}

\title{~~\\ ~~\\ Breaking All the Rules: \\
The Compact Symmetric Object 0402+379}

\author{H. L. Maness\altaffilmark{1,2}, {G. B. Taylor\altaffilmark{2}},
{R. T. Zavala\altaffilmark{2,3}}, {A. B. Peck\altaffilmark{4}}
\& {L. K. Pollack\altaffilmark{5}}}

\altaffiltext{1}{Department of Physics, Grinnell College, Grinnell, IA 50112}
\altaffiltext{2}{National Radio Astronomy Observatory, Socorro NM 87801}
\altaffiltext{3}{Department of Astronomy, New Mexico State University, Las Cruces, NM 88003}
\altaffiltext{4}{Harvard-Smithsonian CfA, SMA Project, PO Box 824, Hilo, HI 96721}
\altaffiltext{5}{Department of Astronomy, University of California at Berkeley, Berkeley, CA 94720}

\slugcomment{As Accepted by the Astrophysical Journal}

\begin{abstract}

  We present results of multi-frequency VLBA observations of the
  compact symmetric object (CSO) 0402+379.  The parsec-scale
  morphology of 0402+379 allows us to confirm it as a CSO, while VLA
  data clearly show the presence of kiloparsec-scale structure.  Thus,
  0402+379 is only the second known CSO to possess large scale
  structure.  Another puzzling morphological characteristic found from
  our observations is the presence of two central, compact,
  flat-spectrum components, which we identify as possible active
  nuclei.  We also present the discovery of neutral hydrogen
  absorption along the southern hotspot of 0402+379 with a central
  velocity $\sim$ 1000 km s$^{-1}$ greater than the systemic velocity.
  Multi-epoch observations from the VLA archive, the Caltech-Jodrell
  Bank Survey, and the VLBA Calibrator Survey allow us to further
  analyze these anomalous features.  Results of this analysis reveal
  significant motion in the northern hotspot, as well as appreciable
  variability in both of the core candidates.  We consider the
  possibility that 0402+379 was formed during a recent merger.  In
  this case, the two candidate cores could be interpreted as binary
  supermassive black holes that have not yet coalesced, whereas the
  large-scale radio emission could be attributed to interactions
  directly linked to the merger or to previous activity associated
  with one of the cores.

\end{abstract}


\section{Introduction}

Compact symmetric objects (CSOs) are a recently identified class of radio
sources smaller than 1 kpc in size with emission on both sides of the 
central engine, and are thought to be very young objects ($\sim$1000 yr, 
Readhead et al. 1996; Owsianik \& Conway 1998).  The small linear sizes
of CSOs make them valuable for studies of both the evolution of radio 
galaxies and for testing the ``unified scheme'' of active galactic
nuclei (AGN).  CSOs are also extremely valuable as calibrators since
they are relatively flux stable, unpolarized, and rarely exhibit any 
extended emission on kiloparsec scales.  For all of these reasons, many
recent observational campaigns have been aimed at identifying and better
understanding these objects (e.g. Peck \& Taylor 2000; Augusto et al. 1998). 
However, despite the increased interest, many properties
of CSOs remain poorly understood.  In particular, the origin of these
objects' unique and varied morphologies remains a 
topic of open debate.  One 
object known to have an especially puzzling morphology is the 
relatively nearby (z=0.055 - Xu et al.~1994) CSO 0402+379.

The radio galaxy 0402+379 was first observed with the Very Long Baseline
Array (VLBA) and the Very Large Array (VLA)\footnote{The 
National Radio Astronomy
Observatory is operated by Associated Universities, Inc., under
cooperative agreement with the National Science Foundation.} 
as part of the first Caltech-Jodrell Bank Survey (CJ1), 
but owing to its moderately large extent (50 mas) and absence of a strong 
compact component, the high resolution image obtained at 5 GHz was quite poor \citep{Xu95}, and
it was not recognized as a CSO.
Complimentary VLA observations obtained during this survey showed 
0402+379 to possess large scale structure, but this result was not surprising
since the CSO morphology was not known.  The  
source was again observed five years later during the VLBA Calibrator 
Survey (VCS; Beasley et al. 2002), but it earned no special attention.  
In an analysis of the polarization properties of core jet sources, 
\citet{Pollack03} classified 0402+379 as a CSO candidate and suggested 
further investigation.  Reexamination of the VCS images
revealed a central core with extended lobes on either
side, consistent with the CSO interpretation.  However, these images
also showed the central core to be offset from a line connecting
the two hotspots, which had not been seen in any other previously identified
CSO. 

In order to directly address the CSO-like morphology of 0402+379, we 
obtained multi-frequency VLBA observations 
at 1.3, 5, and 15 GHz. Our observations allow us to classify this object as a 
CSO.  We emphasize, though, that
0402+379 exhibits several peculiar properties that are not typical of CSOs.
In particular, the presence of large scale structure in this object
is inconsistent with a recent onset of radio activity and has
only been seen in one other CSO \citep[0108+388;][]{Baum90}.  Moreover, the 
parsec-scale structure
of 0402+379 is unique in that it possesses two central, compact, flat-spectrum 
components.  Because jet components are not often found to show these
features, this result is quite puzzling.  

To explore these anomalous features in more detail, we have
re-analyzed multi-epoch data for 0402+379 from the Caltech-Jodrell
Bank Survey (Britzen et al. 2003, in prep.)  and the VLBA Calibrator
Survey \citep{Beasley02}. Our analysis confirms the flat
spectrum nature of this source's two potential nuclei, and
reveals appreciable variability in both of these components.  We also
find significant motion in the northern hotspot of 0402+379 and
possible motion in one of the core candidates.  In \S 4, we explore
possibilities that could account for the two central, compact, flat
spectrum components, as well as the large scale structure and highly
redshifted H{\scriptsize I} discovered in 0402+379.  Throughout this
discussion, we assume H$_{0}$=71 km s$^{-1}$ Mpc$^{-1}$, $\Omega_M$ =
0.27, and $\Omega_{vac}$= 0.73, 
resulting in a linear to angular scale ratio of 1.055 kpc
arcsecond$^{-1}$ \footnote{Derived using E.L. Wright's cosmology
calculator at http://www.astro.ucla.edu/~wright/CosmoCalc.html.}.
    
\section{Observations}

\subsection{2003 VLBA Observations}

VLBA observations were made on 2003 March 02 at 1.348, 4.983, and 
15.353 GHz in a single 11 hr observing session.  A single IF with 
a bandwidth of 16 MHz was observed in 256 channels  
in both R and L circular polarizations, resulting in a frequency resolution 
of 62.5 kHz, corresponding to a velocity resolution of 15 km s$^{-1}$. 
Four-level quantization was employed at all three frequencies.  The net 
integration time on 0402+379 was 651 minutes at 1.3 GHz, 478 minutes at 
5 GHz, and 478 minutes at 15 GHz.  Standard flagging, amplitude calibration, 
fringe fitting, bandpass calibration (using 3C 84 for continuum
data and 3C 111 for spectral line data), and 
frequency averaging procedures were followed in the Astronomical Image 
Processing System (AIPS; van Moorsel et al. 1996).  AIPS reduction scripts
described in \citet{Ulvestad01} were used for a large part of the reduction.
Spectral line Doppler corrections were also applied in AIPS, and a clean cube 
was produced using Difmap \citep{Shepard95}.  
All manual editing, imaging, deconvolution, and 
self-calibration were also done with Difmap.

\subsection{Archival Observations}

We have additionally obtained VLA archival data from 1992 November 
06 and from 1982 July 04 at 1.46 and 4.89 GHz, respectively.  
The 1.46 GHz observation was first presented by \citet{Xu95} and was taken
in the A configuration, whereas the 4.89 GHz observation was taken in the 
BnA configuration and has not been previously published to our knowledge.  
Both frequencies were observed during snapshot
observations, and both were made using a 50 MHz 
bandwidth. Further information regarding these observations can be found 
in Table \ref{Observations}. Standard flagging for these observations as 
well as amplitude and phase calibration was performed in AIPS, while 
manual editing, imaging, deconvolution, and self-calibration were done in Difmap.

To further study this source, we obtained fully-calibrated 
VLBI data taken in 1990 \citep{Xu95}, in 1996 (VCS; Beasley et al. 2002), 
and in three epochs (1994, 1996, and 1999) of the CJ Proper Motion Survey 
(Britzen et al. 2003, in prep.).  These data were 
imaged and modeled in Difmap to aid 
in analysis of motions, variability, and spectra of 0402+379. Further 
information regarding these observations can be found in Table \ref{Observations}.

\section{Results}

\subsection{Radio Continuum}

\begin{figure}
\centering
\vspace{-15mm}
\scalebox{0.3}{\includegraphics{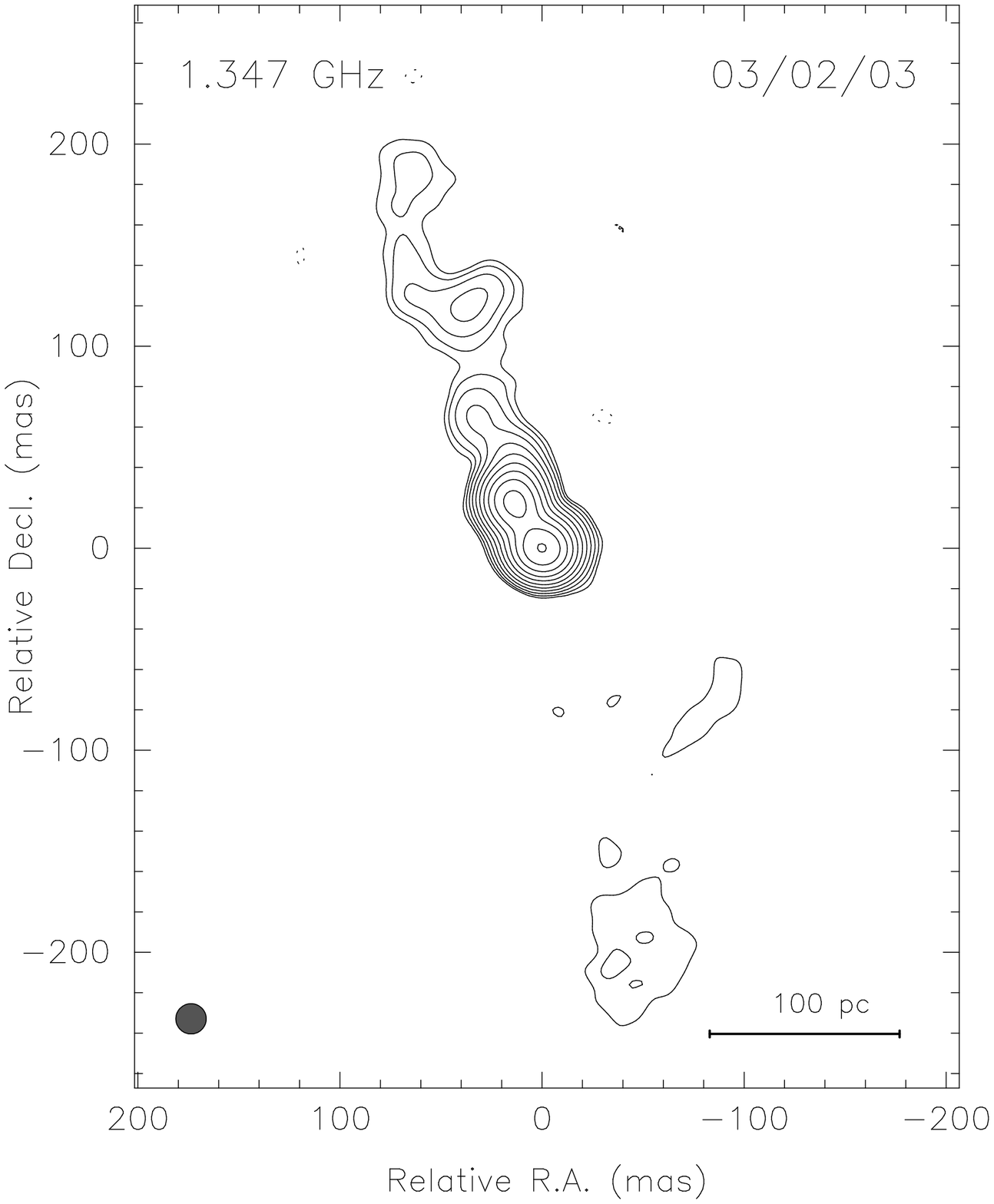}} \\
\vspace{1mm}
\scalebox{0.3}{\includegraphics{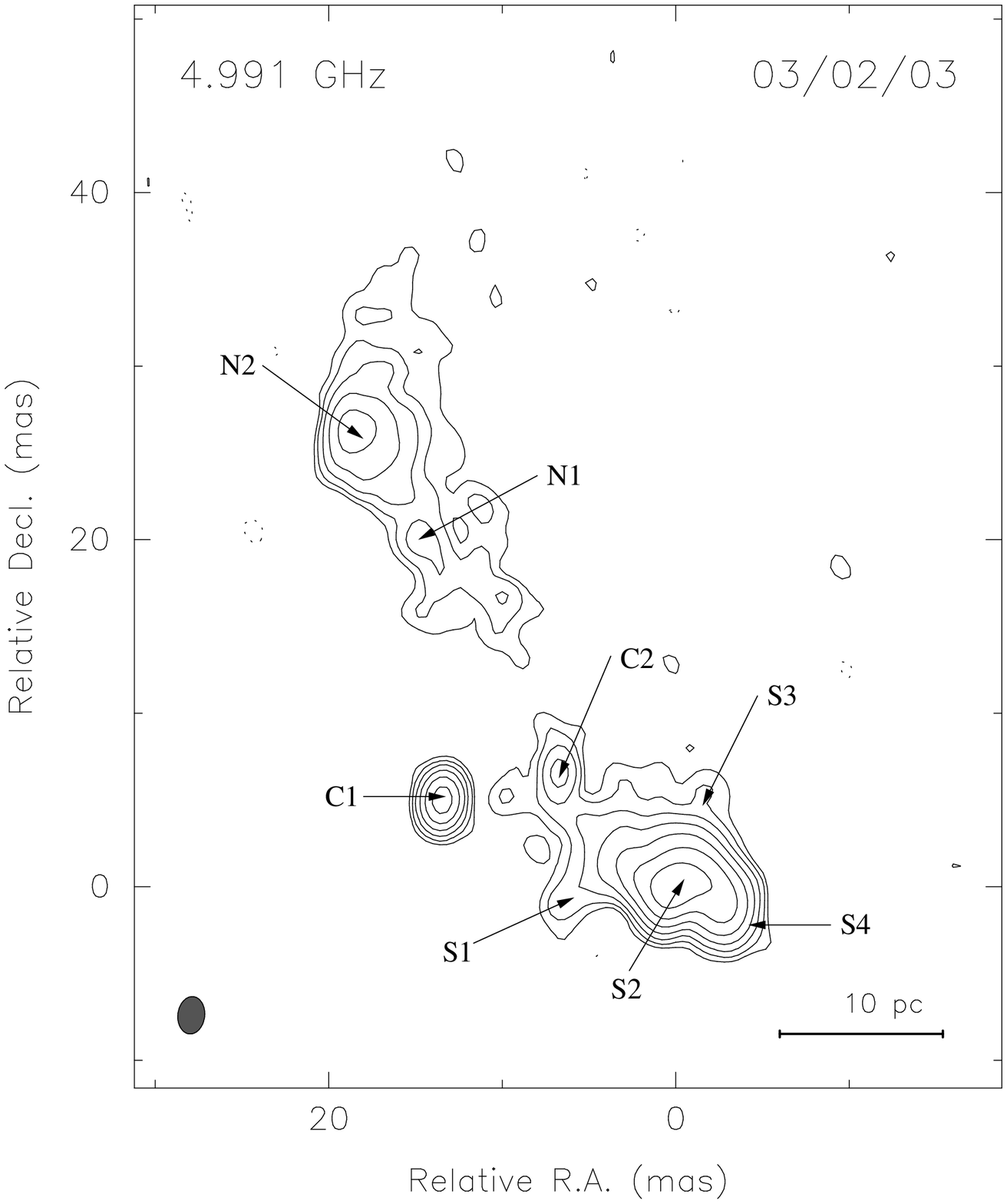}} \\
\vspace{-5mm}
\scalebox{0.3}{\includegraphics{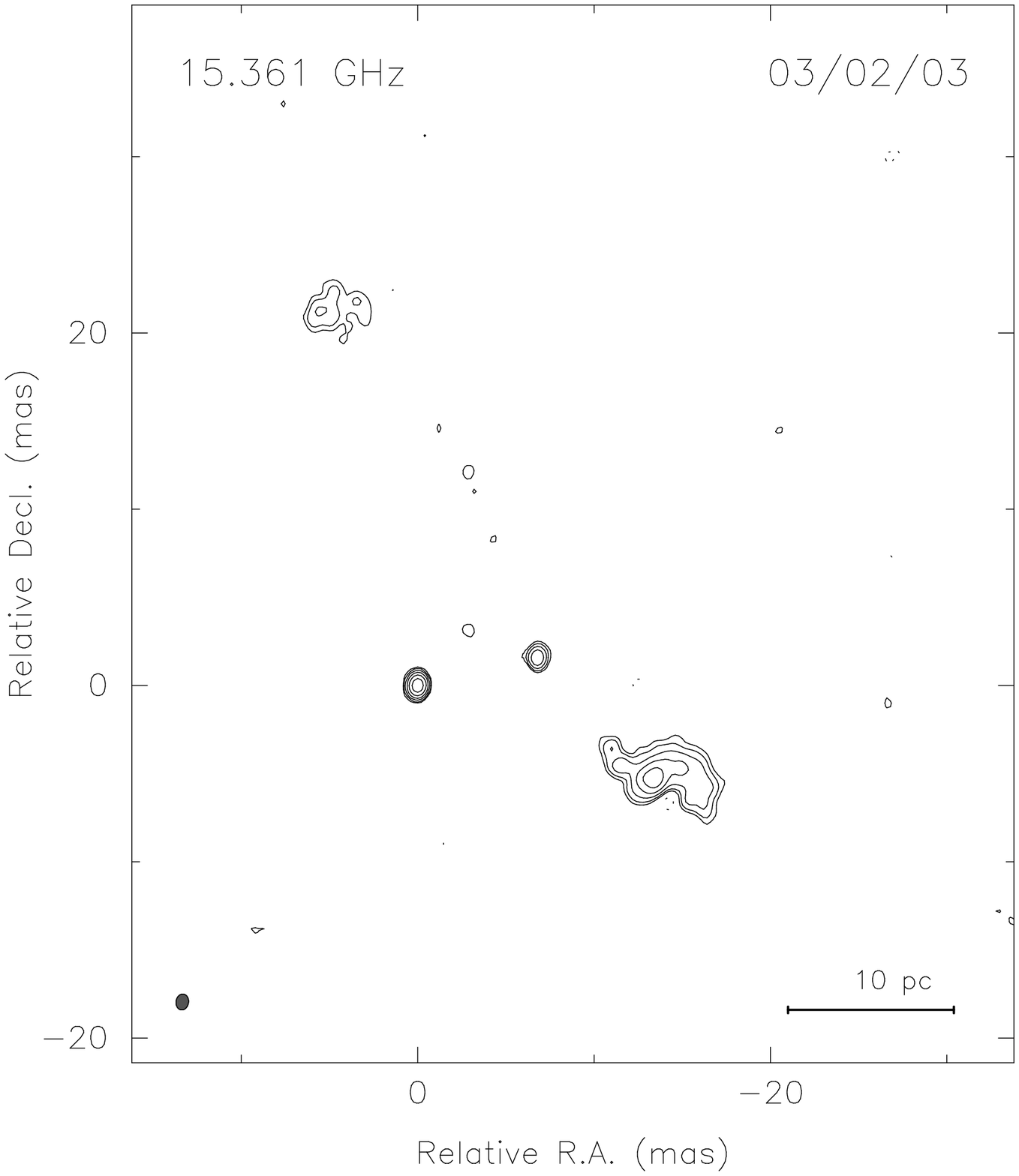}}
\vspace{-1mm}
\caption{Naturally weighted 2003 VLBA images 
of 0402+379 at 1.3, 5, and 15 GHz.   
Contours are drawn beginning at 3$\sigma$ and increase
by factors of 2 thereafter.  Negative contours are dashed.  
The peak flux density and rms noise for each frequency 
are given in Table \ref{Peak_Fluxes}.  The 1.3 GHz map was 
tapered at the FWHM of a circular Gaussian applied to the {\it u,v} model
in order to better show the extended emission in the south.  
The labels shown in the 5 GHz map indicate the 
positions of components derived from model-fitting.
Note the two strong, central compact components.}\label{VLBA03maps}
\end{figure}

Figure \ref{VLBA03maps} 
shows naturally weighted 1.3, 5, and 15 GHz images from the 
2003 VLBA observations.  The overall structure of the source is 
similar at 5 and 15 GHz, spanning $\sim$40 mas ($\sim$40 pc) and 
consisting of two diametrically-opposed jets, as well as two 
strong, compact core candidates, one directly between the jets and 
one also between the jets but offset from the center.  
At 1.3 GHz, the radio morphology is somewhat different, showing  
significant extended structure to the north, as well as weak 
extended structure in the south.  We note that the complex
morphology of the southern hotspot is suggestive
of substantial interaction between the surrounding galactic 
medium and the southern hotspot.

\begin{figure}
\centering
\scalebox{0.4}{\includegraphics{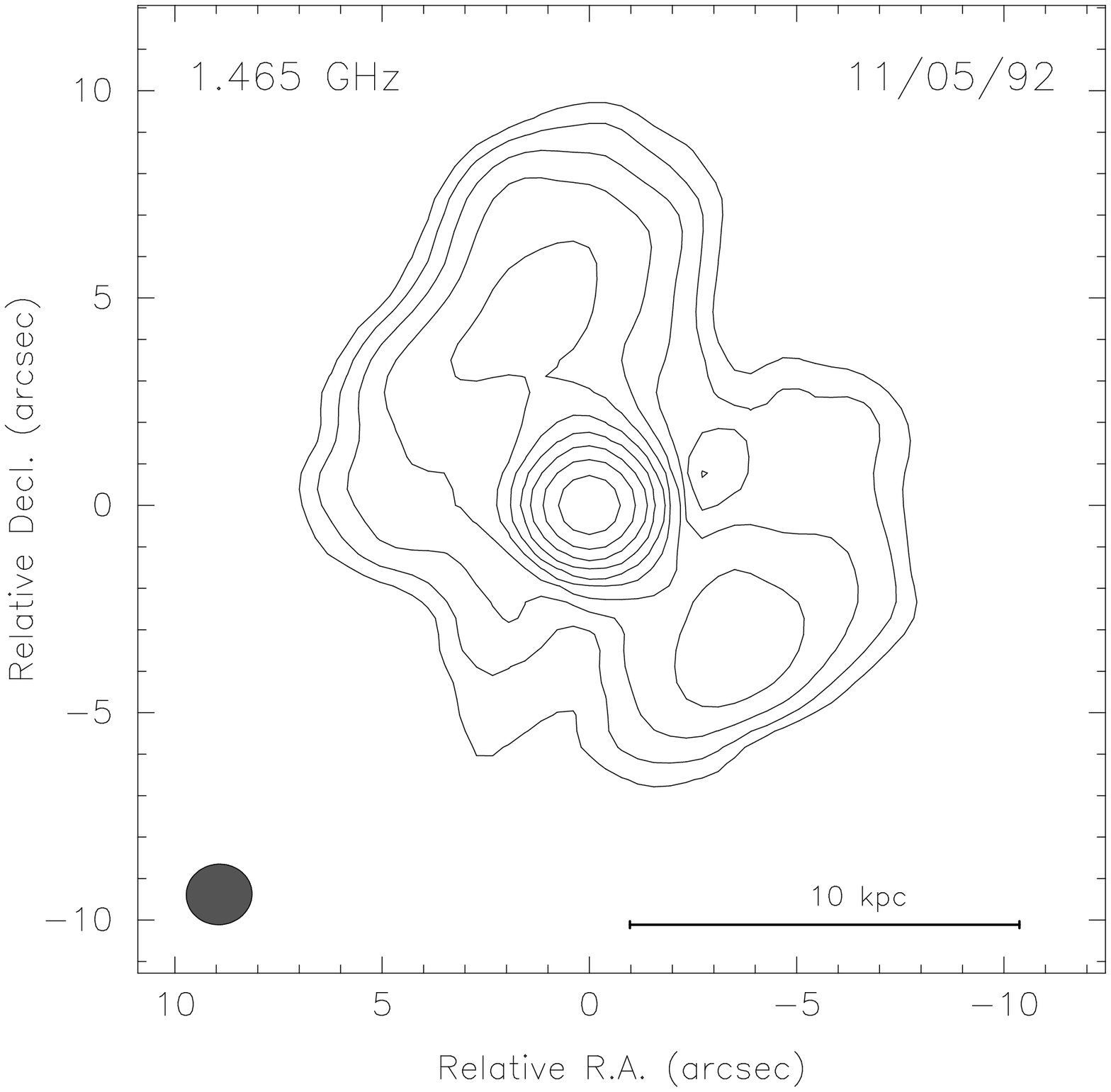}}
\hspace{3mm}
\vspace{10mm}
\scalebox{0.4}{\includegraphics{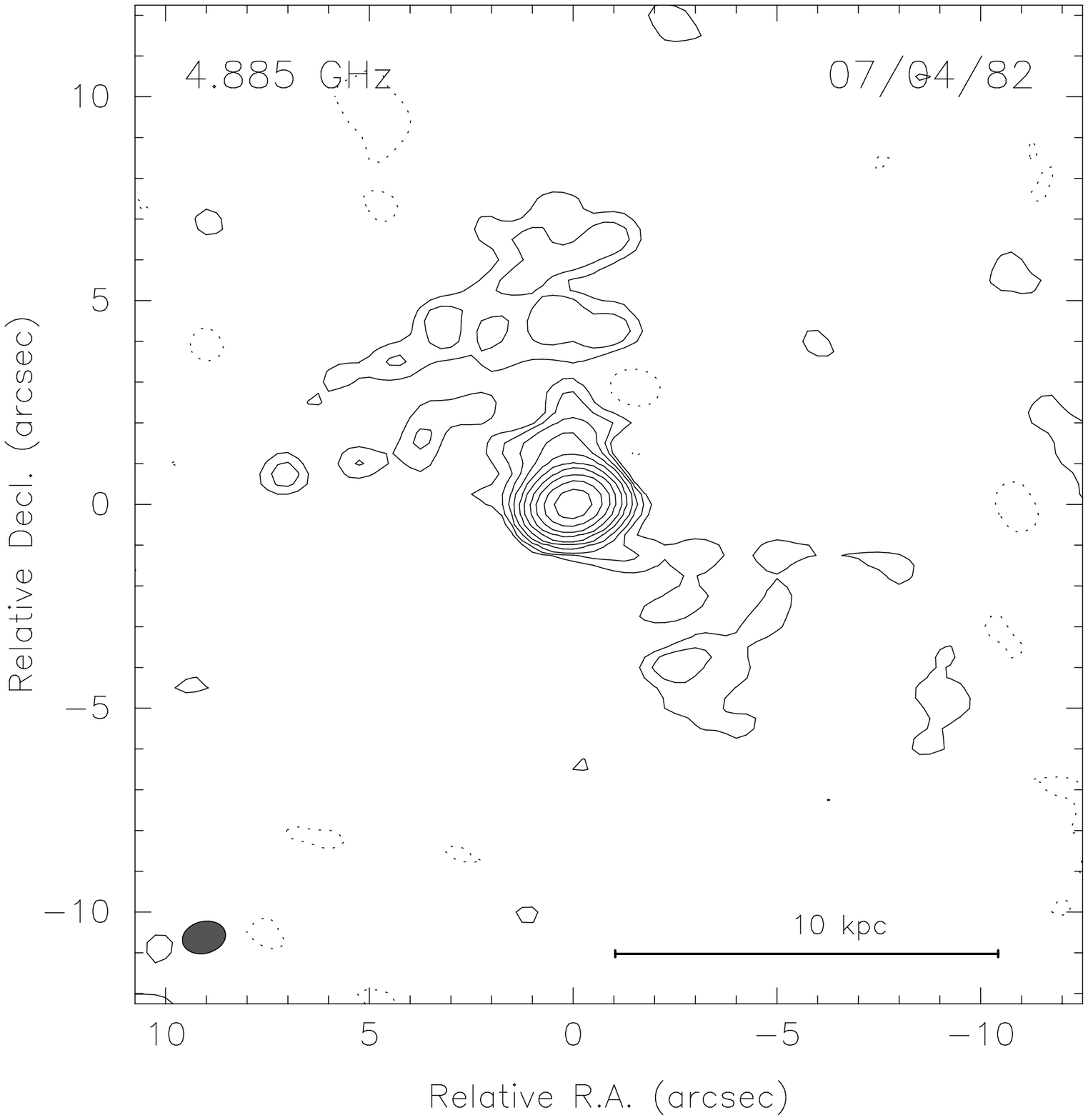}}
\vspace{10mm}
\caption{Uniformly weighted VLA images at 1.4 and 5 GHz.  
The 1.4 GHz data was first presented by Xu et al. (1995), whereas
the 5 GHz data has not been previously published.  Contours are 
drawn beginning at 3$\sigma$ and increase by factors of 2 thereafter.
Negative contours are dashed.  The peak flux density and rms noise are 
1.11 Jy/beam and 0.34 mJy/beam, respectively for the 1.4 GHz map and 
0.94 Jy/beam and 0.53 mJy/beam, respectively for the 5 GHz map.  
Note the asymmetric morphology between the north and south 
lobes in the 5 GHz map.\label{VLAmaps}}
\end{figure}

Results from our VLA observations at 1.4 and 5 GHz are shown in 
Figure \ref{VLAmaps}.  The 1.3 GHz map, initially analyzed by \citet{Xu95}, 
was the first indication of large scale structure in 0402+379.  However, because
the CSO morphology of this source was not known at the time, this
result was not remarkable.  Our VLBA observations, on the other hand,
unambiguously reveal the parsec-scale CSO morphology of this source, thereby
allowing us to report 0402+379 as only the second known CSO (after 0108+388) to possess 
structure on a kiloparsec-scale. 

Because the detection of large scale structure associated with CSOs is
inconsistent with a recent origin for their radio activity,
\citet{Baum90}, who discovered and first analyzed the large scale
structure in 0108+388, have explored possibilities that could account
for extended emission in CSOs.  One possibility they consider is that
the source is a normal radio galaxy in a dense environment, where the
radio plasma cannot currently escape from the nuclear regions.  A
second possibility they give is that the emission in such a source is
recurrent and that the extended radio emission is a relic of activity
from a previous period.  We will return to these possibilities in \S
4.

\subsection{Component Motions and Variability}

In order to explore questions pertaining to motion and variability in
0402+379, we obtained fully-calibrated 5 GHz data from the
Caltech-Jodrell Bank Proper Motion Survey (Britzen et al. 2003, in
prep.).  The three epochs included in this set correspond to VLBA
observations taken in 1994, 1996, and 1999.  Combining this data with
our 2003 VLBA observations and the 1990 VLBI observations of
\citet{Xu95}, we were able to probe motion and variability in this
source over a thirteen year period.  Although the first hot spot
velocities in CSOs required time baselines of $\sim$ 15 years
\citep{Owsianik98}, we note that time baselines as short as 5 years
have also been successfully used to probe motions in CSOs
\citep{Giroletti03}.  Therefore, we believe that our time baseline of
13 years should be sufficient to detect component motions in 0402+379,
should the motion be similar to that in other CSOs.

Motion and variability studies were performed by fitting 8 elliptical
Gaussian components in Difmap to the 1996 visibility data.  The 1996
epoch was chosen for the preliminary fit since data from this epoch
showed a similar morphology to the other epochs but appeared to
be somewhat less complicated in the southern hotspot, making it easier
to fit with a small number of components. In this model, a large
elliptical component was used to pick up the northern jet's diffuse
emission; compact components were used to describe the two core
components, the southern hotspot, and the hotspot in the northern jet.
We used our 1996 model to fit the 5 GHz data corresponding to the
1990, 1994, 1999, and 2003 epochs.  During this process, we let only
position and flux density vary; all other fitting parameters were held
fixed at the 1996 values.  Results from our fits are listed in Table
\ref{Big_Gaussian}, and regions describing each Gaussian component are
labeled in Figure \ref{VLBA03maps}.

\begin{figure}
\centering
\scalebox{0.5}{\includegraphics{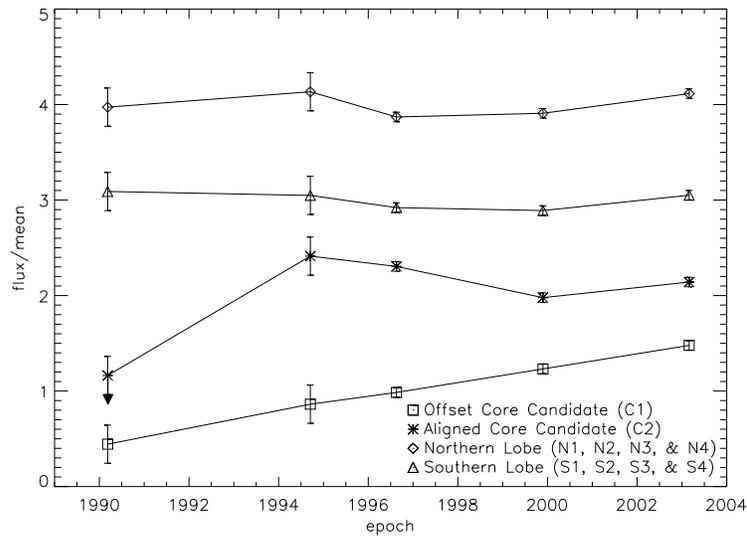}}
\vspace{10mm}
\caption{Lightcurves for selected components at 5 GHz.  The flux densities that
produced this graph were taken from Table \ref{Big_Gaussian} and are discussed 
in \S 3.2. The displayed lightcurves were created by dividing each region's flux
at each frequency by the the mean region flux found from averaging 
all observed frequencies. The offset core candidate, the southern lobe, and the 
northern lobe are displaced 
on the y-axis by 1, 2, and 3 units, respectively.  Errors are estimated from
the rms noise and the absolute flux calibration errors for each epoch.  Note
that component C1 is highly variable, steadily increasing in brightness
over the the thirteen year time baseline, whereas component C2 is not
detected in the 1990 epoch, which suggests that it too is variable.
\label{Lightcurves}}
\end{figure}

To study component variability in 0402+379, we compared the flux for
components C1, C2, the sum of the southern components (S1, S2, S3, and
S4), and the sum of the northern components (N1 and N2) over each of
our five VLBI epochs.  The above regions were chosen based primarily
on their isolation relative to other components in the source.  Errors
for each region were then computed based on the rms noise and our
estimated absolute flux calibration errors ($\sim$ 20\% for the 1990
and 1994 Mk 2 epochs and $\sim$ 5\% for the 1996, 1999, and 2003 VLBA
epochs).  The resulting fractional variation lightcurves are shown in
Figure \ref{Lightcurves}.  These lightcurves were created by dividing
each region's flux at each frequency by the the mean region flux found
from averaging all observed frequencies.  To aid in readability of our
graph, the offset core component (C2), the southern lobe, and the
northern lobe are displaced on the y-axis by 1, 2, and 3 units,
respectively.  From Figure \ref{Lightcurves} and Table
\ref{Big_Gaussian}, we find that component C1 (the offset core
candidate) substantially increases in flux over our thirteen-year
baseline, starting from 18 mJy in 1990 and increasing in brightness to
60 mJy in 2003.  If C1 is the core, we find a relatively high 5 GHz
core fraction of 7\%, compared with the typical CSO 5 GHz core
fraction of $\sim$ 3\% \citep{Taylor96}.  We also find that component
C2 is variable, ranging from $<$10 mJy in 1990 to 25 mJy in 1996.
Because our 1990 epoch was observed with Mk 2 VLBI, this apparent
variability could, in part, be attributed to poor data quality.  We
see, however, that the measured flux for all other components in 1990
is quite consistent with our later epochs, suggesting that our
calculated upper-limit for component C2's flux in 1990 is reliable and
that the observed variability in C2 is significant.  However, this
result is strongly dependent on our 1990 epoch, and further
observations are needed to confirm the variable nature of C2.

Core components in CSOs are known to be variable at $<$23\% on
timescales of several months \citep{Fassnacht01}, significantly less
than both C1 and C2 have proven to be on a timescale of thirteen
years.  Our time sampling is not dense enough to determine the
variability over a period of months in these components, but since the
trend is gradual (with the exception of the change in C2 between 1990
and 1994), our results seem consistent with \citet{Fassnacht01}.  We
therefore believe that the variability we detect in component C1 and
C2 is reasonable for core components in CSOs.  We will further discuss
the implications of these results in \S 4.

\begin{figure}
\centering
\scalebox{0.7}{\includegraphics{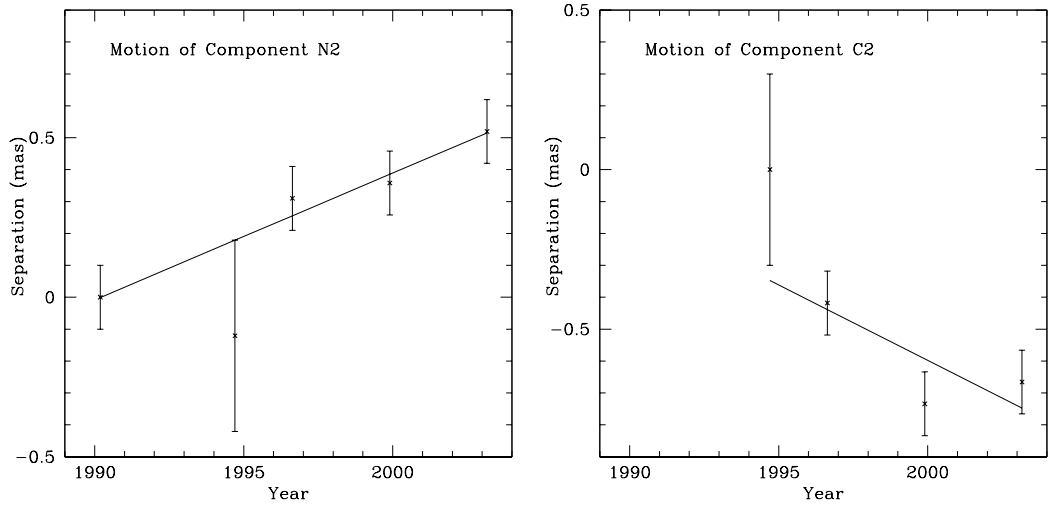}}
\vspace{0mm}
\caption{Plots of relative angular separation versus time (using component C1 as
a reference) for both components in which motion is detected.  
The least-squares velocity fit is overlaid; errors were estimated based
on the scatter exhibited in components for which no motion was detected.  
Note that the velocity fit for component N2 is much better than that
for component C2.\label{motionplot}}
\end{figure}

\begin{figure}
\centering
\scalebox{0.7}{\includegraphics{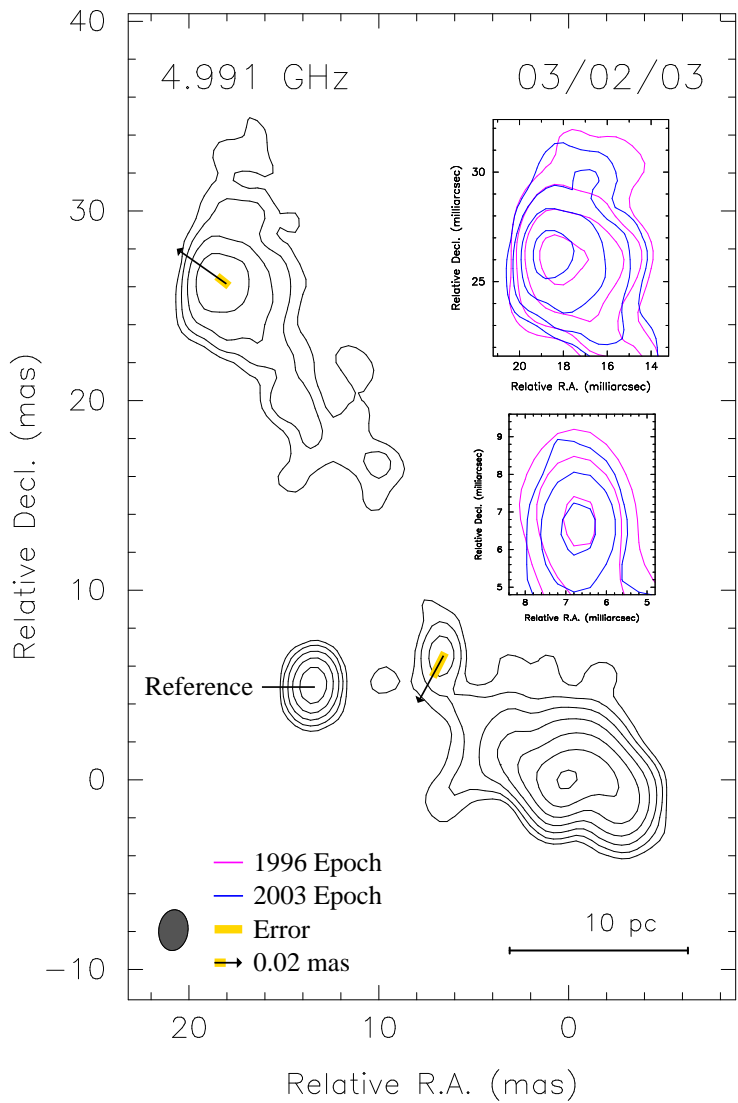}}
\vspace{0mm}
\caption{Schematic diagram of component motions based on the fitting
results presented in Table \ref{Motions_Table}.  The contours are taken
from the 2003 epoch; they start at 5$\sigma$ and increase by factors
of two thereafter.  The red and blue zoomed-in overlays of the moving
components are taken from the 1996 and 2003 epochs, while the overplotted
arrows in the main figure represent the total motion detected over our
thirteen-year time baseline.  Errors are estimated based on the scatter
exhibited in components for which no motion was detected.  Note that we have taken 
component C1 as the reference component for this analysis. \label{motions}}
\end{figure}

To calculate the advance velocity for our Gaussian components, we
chose component C1 as our reference component, based on its isolation
relative to other components in the source, and compared the model-fit
for each epoch by fitting a line to each component's position as a
function of time and then fitting a second line to each component's
velocity as a function of time.  Results of this fitting process are
listed in Table \ref{Motions_Table}, plotted in Figure
\ref{motionplot}, and shown schematically in Figure \ref{motions}.
Position errors at individual epochs were estimated based on the
scatter exhibited in component fits where no motion is detected.
Using our least squares fits shown in Table \ref{Motions_Table}, our
results indicate a $\sim4\sigma$ (0.0398 $\pm$ 0.010 mas yr$^{-1}$ or
0.137 $\pm$ 0.034 c)
detection of motion in the northern hotspot (N2) as well a
$\sim2\sigma$ (0.0468 $\pm$ 0.019 mas yr$^{-1}$ or 0.16 $\pm$ 0.065 c) detection in the
aligned core candidate (C2).  Although it is somewhat surprising that
no motion is detected in the southern hotspot (S2), we emphasize that
the southern hotspot is substantially more complicated than the
northern hotspot (N2) and core candidates (C1 and C2).  Therefore, it
may be that component S2 is moving as fast as our other components
with detected motion (N2 and C2) but that our modelfitting-based
technique does not
allow detection of this motion.
  
Using our velocity result for component N2 given in 
Table \ref{Motions_Table} and our position result for this component 
given in Table \ref{Big_Gaussian}, we 
derive an approximate age for component N2 of 
502 $\pm$ 129 years.  Assuming we have detected
the majority of parsec-scale emission in this source, this component 
age is consistent with that of other expected component ages for CSOs  
(e.g., Readhead et al. 1996; Owsianik \& Conway 1998).

Finally, we emphasize that our detection of motion for component C2 is
extremely tentative, since it is only a $\sim2\sigma$ result.
Moreover, due primarily to the non-detection of component C2 in 1990
as well as the poor quality of data in 1994, the goodness-of-fit for
component C2 is much worse than that for component N2.  We also note
that had we chosen component C2 as stationary, our motion results
would indicate that component N2 is moving somewhat more northwardly
than it is currently moving and at roughly twice its current speed,
whereas component C1 would be moving toward the northern hotspot at
the speed currently attributed to component C2.  We will see in \S4,
however, that the component we choose as a reference is not crucial to
our interpretation, since the only important results are that
component N2 is moving northward and that components C1 and C2 are
possibly moving relative to each other.

\subsection{Radio Continuum Spectra}

\begin{figure}
\centering
\scalebox{0.7}{\includegraphics{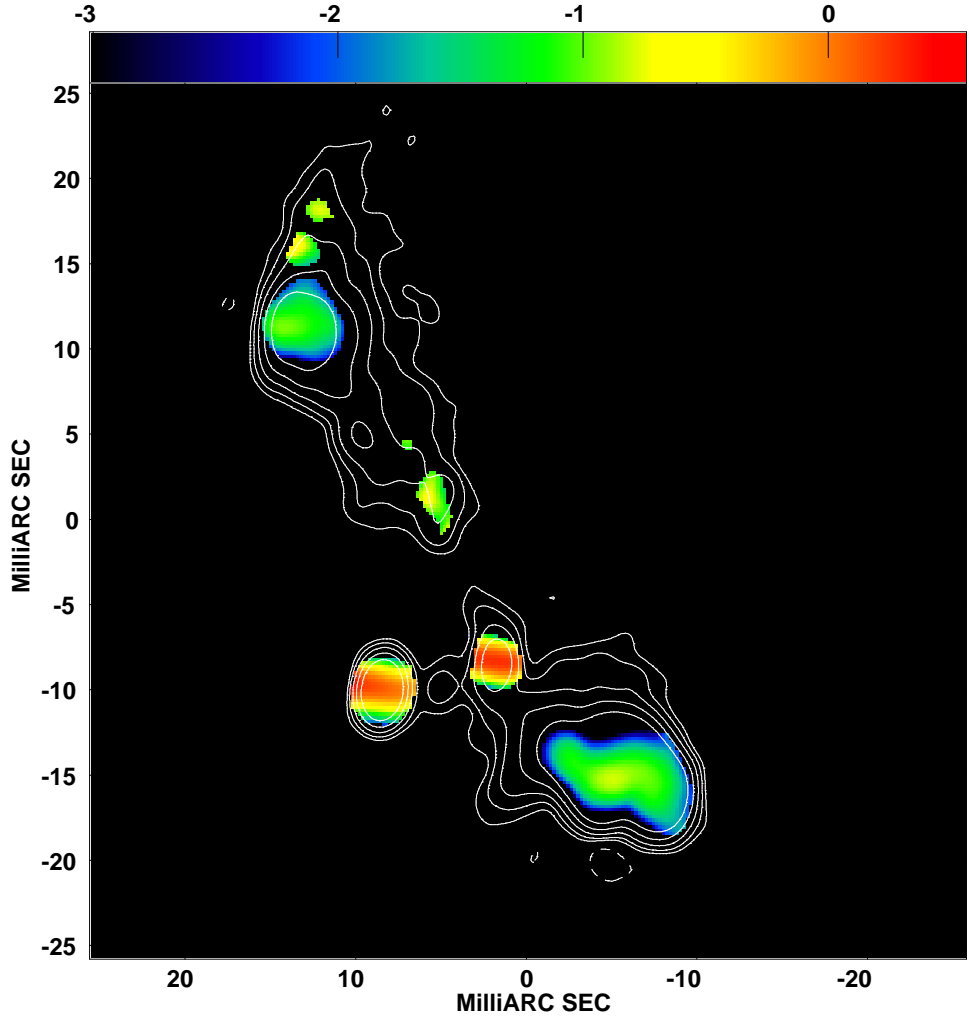}}
\vspace{10mm}
\caption{Spectral index distribution between 5 and 15 GHz
from the 2003 VLBA observations.  The contours are taken from the 5 GHz
observations and are set at 3$\sigma$, increasing by factors of 4 thereafter.
Note that while both hotspots have a steep spectrum,  both core candidates 
(components C1 and C2) have a slightly inverted spectrum. \label{SpecInd}}
\end{figure}

By appropriately tapering our 15 GHz 2003 image, we obtained 
an image resolution matched to our 5 GHz continuum image.  These 
two images were then combined to generate an image of the spectral
index distribution across the source (Figure \ref{SpecInd}).
In both hotspots of the source, a steep spectrum is found, whereas in
both core candidates, the spectrum is quite flat.  Because only core
components in CSOs are generally found to be flat spectrum 
\citep{Taylor96}, we were somewhat surprised by this result. 

\begin{figure}
\centering
\scalebox{0.7}{\includegraphics{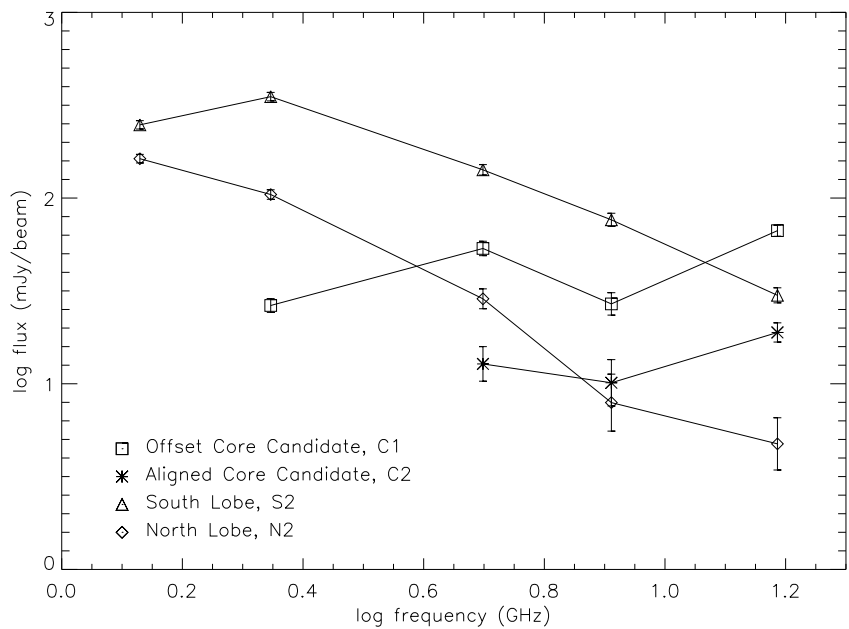}}
\vspace{10mm}
\caption{Continuum spectra for selected components of 0402+379.  The 
spectra were generated from the peak flux measurements given in Table
\ref{Peak_Fluxes}.  Errors are estimated from the rms noise and the absolute
amplitude calibration errors for each epoch.  Note that the aligned core
candidate (C2) is unresolved at 2 GHz and that components C1 and C2 show
a similar spectrum.  We emphasize, though, that the points at 2 and
8 GHz for components C1 and C2 are highly uncertain due to these
components' variability.\label{contspec}}
\end{figure}

To further explore this puzzling feature, we measured the peak flux of
each component using images generated from calibrated data obtained in
the VLBA Calibrator Survey (2 and 8 GHz; 1996 epoch), as well as from
our 2003 epoch at 1.3, 5, and 15 GHz. These flux densities are listed
in Table \ref{Peak_Fluxes}.  Continuum spectra generated from these
flux densities are shown in Figure \ref{contspec}.  We emphasize that
because components C1 and C2 are variable (\S 3.2), our points at 2
and 8 GHz for these components are highly uncertain.  However, the
hotspots in 0402+379 (components N2 and S2) do not appear to be
variable, and we therefore believe that the measured peak fluxes for
these components at 2 and 8 GHz provide a good estimate of the flux we
would have found at these frequencies had we obtained 2003
observations.  We see that our results are consistent with this
supposition since our spectra for components S2 and N2 appear to vary
continuously and because our results obtained for these components are
similar to those obtained from our spectral index map.  Using results
from our continuum spectrum, we find spectral indices between 5 and 15
GHz for the northern hotspot (N2), the southern hotspot (S2), the
offset core candidate (C1), and the aligned core candidate (C2) of
$-$1.38, $-$1.60, 0.20, and 0.35, respectively.

\subsection{H{\scriptsize I} Absorption}

\begin{figure}
\centering
\scalebox{0.8}{\includegraphics{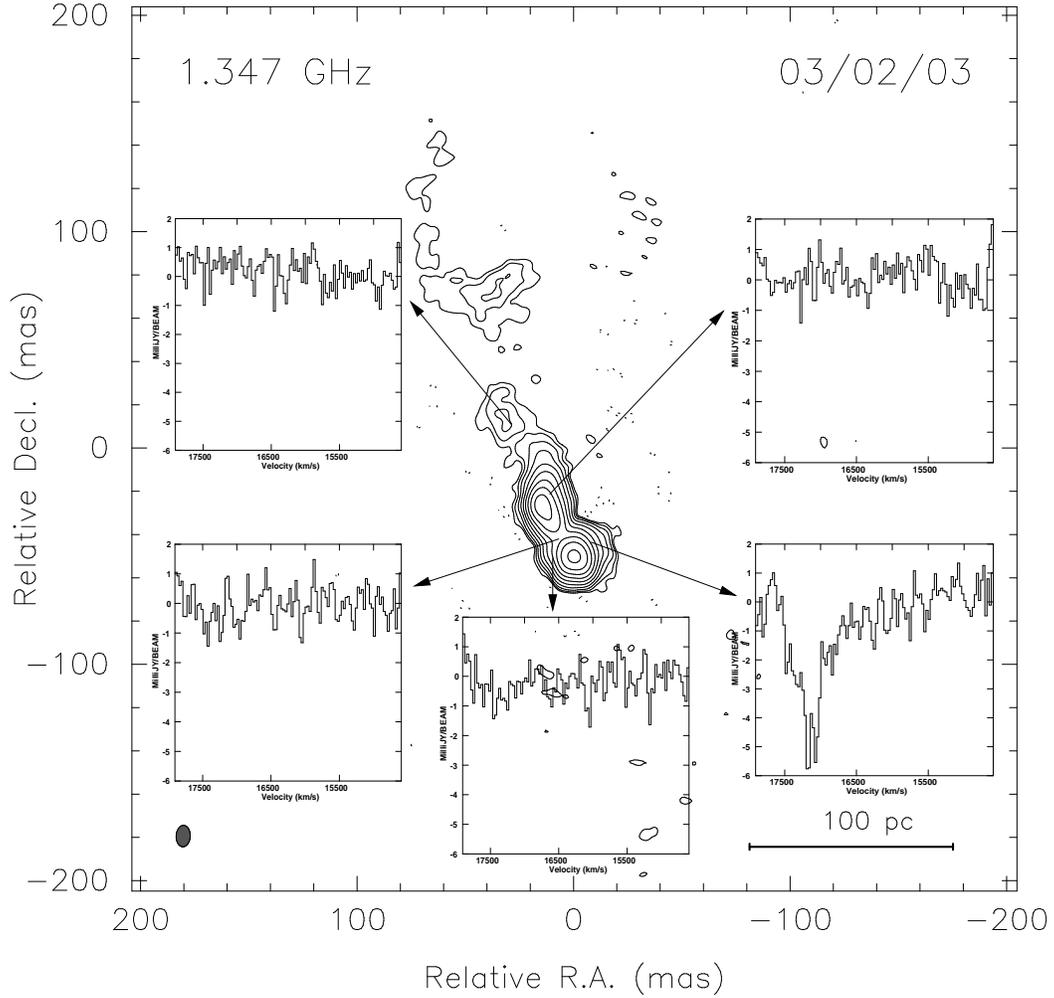}}
\vspace{0mm}
\caption{H{\scriptsize I} absorption profiles taken from five regions of the source.
The systemic velocity is in the center of each spectrum, and the velocity
resolution is 15 km s$^{-1}$.  The continuum has
been subtracted from the spectra in Difmap by removing a continuum model
from the {\it u,v} data of all channels.  The rms noise in a single channel
is 0.55 mJy beam$^{-1}$.\label{HIspec}}
\end{figure}

\begin{figure}
\centering
\scalebox{0.5}{\includegraphics{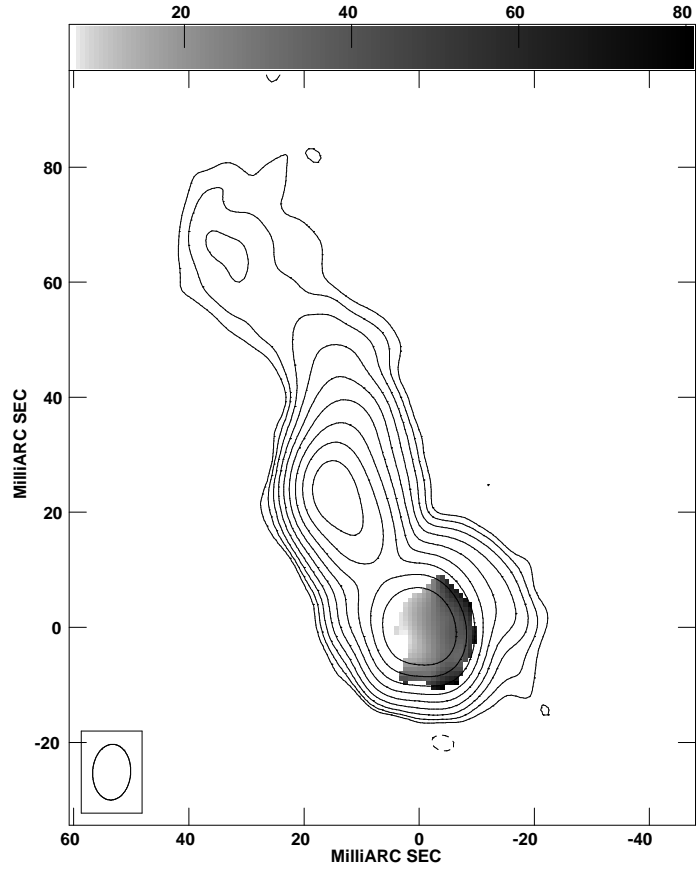}}
\vspace{10mm}
\caption{1.3 GHz continuum map overlaid by the H{\scriptsize I} opacity distribution over
the source.  The map was generated by combining a continuum image of the source
with the continuum-subtracted cube described in Figure \ref{HIspec}.
The channels containing the line were next calculated by fitting Gaussian 
functions at each pixel where the continuum emission is at least 50 mJy.  These 
channels were then summed and blanked at 3$\sigma$ to generate the above map.
\label{Opacity}}
\end{figure}

Spectral line observations at 1.348 GHz were also made during the 2003
VLBA observations. Figure \ref{HIspec} shows the 1.3 GHz continuum
emission from this observation overlaid by spectra from five regions
of the source, while Figure \ref{Opacity} shows a gradient in
H{\scriptsize I} opacity over the source.  A broad (540 km s$^{-1}$)
line is discovered along the eastern edge of the southern hotspot.
Measurements for this line are given in Table \ref{HItab}.  We note
that no velocity gradient was found across the H{\scriptsize I}
absorbing region.

We determine the central velocity of the H{\scriptsize I} absorption
to be 17,115 $\pm$ 23 km s$^{-1}$, corresponding to a redshift of
0.0571 $\pm$ 0.0001.  From the reported redshift for 0402+379, as
measured from optical emission lines \citep{Xu94}, we find that our
observed line is redshifted by 1012 $\pm$ 300 km s$^{-1}$ from the
systemic velocity of the source.  Most of this error comes from the
uncertainty in redshift given by \citet{Xu94}.  Although lines
redshifted greater than 500 km s$^{-1}$ are not typical in CSOs, they
are not unprecedented.  \citet{Vermeulen03}, for example, recently
surveyed 57 low-redshift CSS/GPS sources and found that the majority
of their detected lines were redshifted and were observed at displaced
velocities ranging from several hundred km s$^{-1}$ to sometimes
greater than 1000 km s$^{-1}$.  \citet{Vermeulen03} speculate that in
many of their sources, interaction between the radio jets and the
surrounding galactic medium in the central kpc lead to
significant motions in the H{\scriptsize I} absorbing gas and, thus,
result in absorption lines that are greatly displaced from the
systemic velocity.  This hypothesis was also used to explain the
highly blueshifted line found in IC~5063 \citep{Oosterloo00}.  Because
the morphology of 0402+379 in the H{\scriptsize I} absorbing region of
the southern hotspot is highly suggestive of substantial interaction
between the radio jets and the galactic medium (as previously
discussed in \S 3.1), a similar explanation could be used to explain
our substantially redshifted line.  We note that while this hypothesis
may appear to be in conflict with our nondetection of motion in the
southern hotspot (\S 3.2), as we have already discussed, component S2
is too complicated for proper motions less than 0.3c to be detected.
We further note that assuming this hypothesis is correct may also
suggest that the southern jet is on the receding side with respect to
our line-of-sight, since the absorption line is redshifted from the
systemic velocity.  This is consistent with the considerably weaker
emission in the southern extension of the source than in the northern
extension in the continuum maps at 1.3 GHz with the VLBA and at 5 GHz
with the VLA.  A problem with this hypothesis is how to sweep up
enough neutral material in the shock driven into the cloud
\citep{Bicknell98}.

As an alternative hypothesis, it may be possible that
the highly redshifted H{\scriptsize I} in 0402+379 is due to a
recent merger.  In this case, the observed H{\scriptsize I} redshift
of $\sim$ 1000 km s$^{-1}$ would be attributed to the velocity
difference between the two merging galaxies, or to infall of gas that
has been dispersed during the merger.  

\section{Discussion}

Although the clearly symmetric morphology of the lobes in 0402+379
allows us to classify it as a CSO, 0402+379 exhibits several
peculiar properties that are not typical of CSOs.  In particular,
0402+379 has been shown to have large scale structure, making it 
only the second known
CSO to possess such a feature.  In addition, 0402+379 contains
two central, compact, flat spectrum, variable components, a feature which
has not been observed in any other compact source.  While we are unable
to fully explain these anomalous properties at present, in the following
sections, we discuss several physical situations that could account for one or 
more of the observed features.

\subsection{Background Source}

To explain the two central compact, flat spectrum components observed
in VLBI images of 0402+379, we first considered the possibility that
component C1 is not actually associated with 0402+379 but is instead a
foreground or background radio source.  However, from our angular
separation distance for the two core candidates (7 mas), we find the
probability that component C1 is an unassociated background source to
be $\sim$10$^{-11} $\citep{Condon84}.  Moreover, our 5 GHz map shows a
6$\sigma$ detection of radio emission connecting components C1 and C2
(e.g. Figure \ref{SpecInd}).  We therefore eliminate the background
source theory as a likely explanation for the two central compact,
flat spectrum components observed in 0402+379.

\subsection{Gravitational Lensing}

A second possibility that could account for the two, central compact
components we observe in 0402+379 is that our source is
gravitationally lensed.  Starting in the late 1980's, a number of
gravitationally-lensed radio sources have been observed and documented
(e.g., Soucail et al. 1987).  The smallest known lens is 0218+357,
with a separation of 330 mas \citep{Odea92}. In response to the
moderate lens separations of \citet{Odea92} and others,
\citet{Wilkinson01} surveyed three hundred compact radio sources in
search of lensing (with an expected lensing mass of $\sim$$10^7{\rm
  M}_{\Sun}$) and found none that were gravitationally lensed.
Therefore, if 0402+379 were lensed, it would be the first in its class
to be identified as such.  Moreover, we note that while the morphology
and spectrum of 0402+379 are highly supportive of the lensing
hypothesis, we see that the lightcurves of components C1 and C2 (the
two core candidates) are very different (see Figure
\ref{Lightcurves}).  Although the lightcurves for a lensed source are
not required to be identical, we would expect them to be proportional,
with a short time delay.  In addition, because 0402+379 is nearby
(z=0.055), the lensing mass would likely be obvious if 0402+379 were
lensed. Therefore, based on the known rarity of compact, lensed
sources, the significant difference in the lightcurves of components
C1 and C2, and the close proximity of 0402+379, we eliminate lensing
as a potential mechanism to describe the observed parsec-scale
morphology.

\subsection{Dense Medium Theory}

First analyzed by \citet{Baum90}, 0108+388 was the first CSO found to
possess large scale structure.  Because the presence of kpc-scale
emission associated with CSOs is inconsistent with a recent origin for
their radio activity, \citet{Baum90} have presented several theories
that could reconcile this apparent contradiction.  One possibility
they present is that 0108+388 is a CSO\footnote{\citet{Baum90}
actually referred to 0108+388 as a compact double (CD) since the
classification of CSO was not in existence at the time of their work.
However, 0108+388 has since been classified as a CSO
\citep{Readhead96}, so in all discussions of this source, we will
refer to it as such.}, but that the radio emission is not a recent
event.  \citet{Baum90} speculate, then, that 0108+388 could be a more
evolved galaxy but that the radio plasma cannot currently escape from
the nuclear regions.  They propose that this situation might arise
because the host galaxy has recently swallowed a gas rich companion,
thereby smothering the source.  Since 0402+379 is similar to 0108+388
as a CSO with large scale structure, we consider this possibility in
some detail.

Because exotic jet morphologies are often attributed to substantial
interaction with the external medium, the medium-interaction theory
presented by \citet{Baum90} could explain the puzzling pc-scale
emission seen in 0402+379.  In this case, the offset-core candidate
(component C1) would be classified as the core, whereas the
aligned-core candidate (component C2) would be classified as a knot in
the southern jet.  The southern hotspot morphology, the presence of
greatly redshifted H{\scriptsize I}, and the lack of significant
motion in the southern hotspot all appear to support this theory,
since they all suggest substantial interaction between the ISM and the
southern hotspot. Moreover, although the inverted spectrum of
component C2 is difficult to explain using our ``dense medium''
hypothesis, the non-detection of component C2 in 1990 could suggest
that substantial interactions with a dense surrounding medium have led
to shocks in component C2, causing it to brighten significantly, as
well as to appear compact and flat-spectrum.  However, our derived
H{\scriptsize I} column density of 1.8 $\times 10^{20}$ cm$^{-2}$ in
the southern hotspot is inconsistent with the slow jet, dense medium
model for CSOs presented by \citet{Readhead96} by three orders of
magnitude. It could be argued that this discrepency is due to our
assumed spin temperature of 100 K, which was used in deriving the
column density.  We note, though, that while this assumed temperature
may be underestimated, the column density derived assuming a spin
temperature of 10$^4$ K is still too small to be consistent with the
dense medium model presented by \citet{Readhead96}. If our dense medium
theory is correct, then our H{\scriptsize I} opacity results require
an anomalously large density gradient between component C2 and the
southern hotspot.  Our detection of significant motion in the northern
hotspot, as well as the fact that we see no morphological evidence for
substantial interaction between the northern jet and the surrounding
medium also weakens our dense medium theory, since we would expect the
medium to be distributed isotropically about the source.  The relative
motion of $\sim$0.16 c between C1 and C2 is plausible if C2 is a knot
moving down a curved jet.  Finally, as \citet{Baum90} suggest, if the
parsec-scale structure in our source is smothered by a dense medium,
we might suppose that this density arises due to a merger between the
host galaxy and a gas rich companion.  If this were the case, we might
expect to see some evidence for the interaction such as a disturbed
optical morphology, or peculiar optical emission lines (e.g., Bartel
et al. 1984).  However, 0402+379 has a typical LINER spectrum
\citep{Ho97} and does not contain any unusual features
\citep{Stickel93}.  The digital sky survey images show nothing unusual
in the bright elliptical host galaxy with a magnitude of 17.2
\citep{Xu94}.

\subsection{Supermassive Binary Black Holes}

A final possibility that could account for several of the observed
features in 0402+379 is that the two observed flat-spectrum, compact
components are part of a supermassive black hole binary system formed
from a recent merger.  In this case, components C1 and C2 (the two
core candidates) would both be interpreted as active nuclei, with the
northern and southern hotspots belonging to component C2 (the aligned
core candidate).  The morphology and the spectrum are highly supportive
of this hypothesis.   In this
picture, the high velocity of the H{\scriptsize I} absorption could be
attributed to infalling gas from a merger.

\citet{Readhead96} remarked on the commonality of an ``S-shaped''
morphology
in CSOs and suggested that it could be the result of precession
due to a binary black hole system in the nucleus.  
We note that 0108+388, the only other CSO known to have large scale
structure, has S-symmetry.  Because the
extended strucuture we see in 0108+388 and in 0402+379 is quite
anomalous in CSOs, it is reasonable to suppose that these objects are
part of similar systems and were formed via similar mechanisms.
\citet{Baum90} have suggested that the large scale structure observed
in 0108+388 is a relic of a previous period of activity.  Given the
possibility that both 0108+388 and 0402+379 harbor binary black holes
in their centers, the previous period of activity may have been linked
to a recent merger. We note, however, that we do not see an S-shaped
morphology in the jets of 0402+379, which detracts from this
argument.

Although the claim by \citet{Readhead96} that the S-shaped
morphologies in the jets of CSOs is indicative of a binary black hole
system was speculative, more recent studies have provided substantial
evidence that such systems do exist.  For example, \citet{Sudou03}
have recently used high precision astrometry with the VLBA to 
observe an elliptical
orbit in the core of the radio galaxy 3C 66B, providing strong
evidence for a supermassive black hole companion.  Moreover, it has
been theorized that a small fraction of bright elliptical galaxies
should harbor binary black holes at their center \citep{Haehnelt02}.
It is not surprising that binary systems with a linear separation on
the same scale at which we observe the core candidates in 0402+379
have not previously been imaged, since these objects are likely quite
rare and since it is not required that either or both of the black
holes be active.

We note, however, that our 2$\sigma$ result that component C2 is
moving relative to component C1 may weaken our binary black hole
hypothesis.  Assuming typical supermassive black hole masses of
$\sim$$10^8M_{\Sun}$ for the binary system and using the radial
separation derived from our 2003 maps (7 pc), we find from Kepler's
laws that the period of rotation for a binary supermassive black hole
system such as ours should be on the on the order of $\sim$$10^4$
years.  This period corresponds to an orbital velocity between
components C1 and C2 of $\sim$0.001c, which is two orders of magnitude
slower than our measured relative velocity of $\sim$0.16 c (see
Table \ref{Motions_Table}).  Given the total merger time expected for
galactic collisions \citep{Haehnelt02}, it is improbable that we would
observe binary black holes in 0402+379 at the measured separation
distance between components C1 and C2 if the black holes were not in a
stable orbit.  We emphasize, though, that the above calculation
assumes a stable, Keplerian orbit, which might be a poor assumption.
Moreover, the detection of relative motion between components C1 and
C2 is only a 2$\sigma$ result, so C1 and C2 may not be moving
significantly with respect to each other.  We therefore believe that
our motion results do not yet rule out the binary black holes explanation.

\section{Conclusion}

We have identified two plausible explanations for the anomalous
features observed in 0402+379: (1) interactions with an unusually
dense and complex circumnuclear medium; and (2) a binary black hole
system.  However, we find some features from our observations that
neither theory can adequately explain.  Namely, from our continuum
maps, our motion results, and our H{\scriptsize I} measurements, we
find that the medium surrounding 0402+379 does not appear to be dense
enough to support the ``dense medium theory.''  In addition, the
relative motion we detect between our two core candidates may be too
great to support a theory that both central, compact, flat-spectrum
components in our source are binary active nuclei, since the
probability of detecting the system in an unstable orbit is quite low.

Because 0402+379 possesses several unusual, anomalous features,
constraining the theories we use to describe it is absolutely
necessary if we are to better understand CSOs in the unified scheme of
AGN.  Future observations could significantly help in this process.
Obtaining another epoch in 2005 at 5 and 15 GHz would allow us to
better probe the motion in 0402+379 and potentially allow us to test
the supermassive binary black holes theory.  Low frequency (i.e. 90
cm) VLBA observations at 50 mas resolution would also be helpful
because they would allow us to search for a connection between our
current VLA and VLBA observations, thereby providing more information
regarding the large scale structure in 0402+379.  In addition, high
frequency VLBA observations at 22 and 43 GHz would allow us to search
for a spectral turnover in the two core candidates, which would be
very helpful in identifying the true core in this object if only one
of our components is a nucleus.  Finally, multi-wavelength
observations could constrain theories describing this object. Although
the host galaxy of 0402+379 is fairly bright with a magnitude of 17.2
\citep{Xu94}, it has not been imaged in detail.  Particularly useful,
then, would be to image this source at high resolution in order to
look for optical evidence of a merger or disturbances in the nucleus.

\acknowledgments{The authors are grateful to S. Britzen and
R.C. Vermeulen for supplying the 1994, 1996, and 1999 data in advance
of publication.  H.M. is also thankful to Ylva Pihlstr\"om for her
guidance and help on this project and to the REU program at NSF and
the National Radio Astronomy Observatory for funding this research.
In addition, R.T.Z. gratefully acknowledges support from a predoctoral
fellowship from NRAO and from the New Mexico Alliance for Graduate
Education and the Professiorate through NSF grant HRD-0086701,
and A.B.P. thanks NRAO for hospitality during part of this
project.  The authors are grateful to an anonymous referee for several 
insightful suggestions. }

\clearpage

\begin{deluxetable}{lccccccc}
\tabletypesize{\scriptsize}
\tablecolumns{8}
\tablewidth{0pt}
\tablecaption{Observations\label{Observations}}
\tablehead{\colhead{Frequency}&\colhead{Instrument}&\colhead{Date}&\colhead{Integ. Time}&\colhead{BW}&\colhead{Polar.}&\colhead{IFs}&\colhead{Reference}\\
\colhead{(GHz)}&\colhead{}&\colhead{}&\colhead{(min)}&\colhead{(MHz)}&\colhead{}&\colhead{}&\colhead{}}
\startdata
1.39 & VLBA & 02 Mar 2003 & 651 & 16 & 2 & 1 & This Paper \\
1.46 & VLA & 06 Nov 1992 & 4 & 50 & 2 & 2 & Xu et al. 1995 \\
2.22 & VLBA & 06 July 1996 & 3 & 8 & 1 & 4 & Beasley et al. 2002 \\  
4.89 & VLA & 04 July 1982 & 2 & 50 & 2 & 1 & Unpublished \\
4.99 & VLBI Mk 2 & 10 Mar 1990 & 80 & 2 & 1 & 1 & Xu et al. 1995 \\
4.99 & VLBI Mk 2 & 17 Sep 1994 & 57 & 2 & 1 & 1 & Britzen et al. 2003, in prep. \\
4.99 & VLBI & 19 Aug 1996 & 41 & 8 & 1 & 1 & Britzen et al. 2003, in prep. \\
4.99 & VLBA & 26 Nov 1999 & 35 & 8 & 2 & 2 & Britzen et al. 2003, in prep. \\
5.00 & VLBA & 02 Mar 2003 & 478 & 16 & 2 & 1 & This Paper \\
8.15 & VLBA & 09 July 1996 & 3 & 8 & 1 & 4 & Beasley et al. 2002 \\
15.35 & VLBA & 02 Mar 2003 & 478 & 16 & 2 & 1 & This Paper \\
\enddata
\end{deluxetable}

\begin{deluxetable}{lcccccccc}
\tabletypesize{\scriptsize}
\tablecolumns{9}
\tablewidth{0pt}
\tablecaption{Gaussian Model Components\tablenotemark{*}.\label{Big_Gaussian}}
\tablehead{\colhead{Component}&\colhead{Epoch}&\colhead{S}&\colhead{r}
&\colhead{$\theta$}&\colhead{a}&\colhead{b/a}&\colhead{$\Phi$}
&\colhead{$\chi^{2}$} \\
\colhead{} & \colhead{} & \colhead{(Jy)} & \colhead{(mas)} 
& \colhead{($^o$)}&\colhead{(mas)}&\colhead{}&\colhead{($^o$)}}
\startdata
C1... & 1990 & 0.018 $\pm$ 0.004 &  0.0   &     0.0 & 0.525 & 1.00 &  172.4  & 1.13 \\
      & 1994 & 0.040 $\pm$ 0.009 &  0.0   &     0.0 & 0.525 & 1.00 &  172.4  & 0.90 \\
      & 1996 & 0.040 $\pm$ 0.003 &  0.0   &     0.0 & 0.525 & 1.00 &  172.4  & 1.31 \\  
      & 1999 & 0.050 $\pm$ 0.004 &  0.0   &     0.0 & 0.525 & 1.00 &  172.4  & 1.28 \\
      & 2003 & 0.060 $\pm$ 0.004 &  0.0   &     0.0 & 0.525 & 1.00 &  172.4  & 2.07 \\
C2... & 1990 & $<$0.010 &     &         &       &      &         &      \\
      & 1994 & 0.025 $\pm$ 0.006 &  7.323 &  -73.79 & 1.61  & 1.00 &  -152.1 & 0.90 \\
      & 1996 & 0.024 $\pm$ 0.002 &  6.932 &  -75.42 & 1.61  & 1.00 &  -152.1 & 1.31 \\
      & 1999 & 0.018 $\pm$ 0.001 &  6.803 &  -78.00 & 1.61  & 1.00 &  -152.1 & 1.28 \\ 
      & 2003 & 0.021 $\pm$ 0.001 &  6.809 &  -77.28 & 1.61  & 1.00 &  -152.1 & 2.07 \\
S1... & 1990 & 0.078 $\pm$ 0.017 & 11.063 & -105.26 & 3.29  & 1.00 &  -51.7  & 1.13 \\
      & 1994 & 0.090 $\pm$ 0.020 & 12.154 & -108.78 & 3.29  & 1.00 &  -51.7  & 0.90 \\
      & 1996 & 0.087 $\pm$ 0.006 & 11.978 & -108.54 & 3.29  & 1.00 &  -51.7  & 1.31 \\
      & 1999 & 0.090 $\pm$ 0.006 & 11.600 & -108.17 & 3.29  & 1.00 &  -51.7  & 1.28 \\
      & 2003 & 0.119 $\pm$ 0.007 & 11.846 & -108.81 & 3.29  & 1.00 &  -51.7  & 2.07 \\
S2... & 1990 & 0.291 $\pm$ 0.064 & 14.119 & -111.55 & 1.51  & 1.00 &  -57.7  & 1.13 \\
      & 1994 & 0.228 $\pm$ 0.050 & 14.364 & -111.54 & 1.51  & 1.00 &  -57.7  & 0.90 \\
      & 1996 & 0.211 $\pm$ 0.015 & 14.234 & -111.29 & 1.51  & 1.00 &  -57.7  & 1.31 \\ 
      & 1999 & 0.198 $\pm$ 0.014 & 14.120 & -111.34 & 1.51  & 1.00 &  -57.7  & 1.28 \\
      & 2003 & 0.208 $\pm$ 0.012 & 14.165 & -111.12 & 1.51  & 1.00 &  -57.7  & 2.07 \\
S3... & 1990 & 0.187 $\pm$ 0.041 & 16.029 & -106.60 & 3.40  & 0.37 &   32.0  & 1.13 \\
      & 1994 & 0.168 $\pm$ 0.037 & 16.144 & -107.25 & 3.40  & 0.37 &   32.0  & 0.90 \\ 
      & 1996 & 0.158 $\pm$ 0.011 & 16.149 & -107.52 & 3.40  & 0.37 &   32.0  & 1.31 \\
      & 1999 & 0.152 $\pm$ 0.011 & 16.084 & -107.61 & 3.40  & 0.37 &   32.0  & 1.28 \\
      & 2003 & 0.186 $\pm$ 0.011 & 16.015 & -107.52 & 3.40  & 0.37 &   32.0  & 2.07 \\
S4... & 1990 & 0.006 $\pm$ 0.001 & 17.454 & -108.73 & 2.72  & 0.10 &  -26.2  & 1.13 \\
      & 1994 & 0.028 $\pm$ 0.006 & 17.399 & -107.40 & 2.72  & 0.10 &  -26.2  & 0.90 \\
      & 1996 & 0.019 $\pm$ 0.001 & 17.731 & -109.84 & 2.72  & 0.10 &  -26.2  & 1.31 \\
      & 1999 & 0.019 $\pm$ 0.001 & 17.641 & -111.04 & 2.72  & 0.10 &  -26.2  & 1.28 \\
      & 2003 & 0.029 $\pm$ 0.002 & 17.763 & -110.29 & 2.72  & 0.10 &  -26.2  & 2.07 \\
N1... & 1990 & 0.107 $\pm$ 0.024 & 17.033 &    5.15 & 13.58 & 0.27 &   20.3  & 1.13 \\ 
      & 1994 & 0.093 $\pm$ 0.020 & 18.678 &    4.72 & 13.58 & 0.27 &   20.3  & 0.90 \\
      & 1996 & 0.086 $\pm$ 0.006 & 19.143 &    5.95 & 13.58 & 0.27 &   20.3  & 1.31 \\
      & 1999 & 0.085 $\pm$ 0.006 & 18.874 &    5.45 & 13.58 & 0.27 &   20.3  & 1.28 \\
      & 2003 & 0.105 $\pm$ 0.006 & 19.271 &    6.13 & 13.58 & 0.27 &   20.3  & 2.07 \\
N2... & 1990 & 0.044 $\pm$ 0.010 & 21.103 &   12.86 &  2.38 & 1.00 & -105.8  & 1.13 \\ 
      & 1994 & 0.057 $\pm$ 0.013 & 21.414 &   11.47 &  2.38 & 1.00 & -105.8  & 0.90 \\   
      & 1996 & 0.049 $\pm$ 0.003 & 21.416 &   13.17 &  2.38 & 1.00 & -105.8  & 1.31 \\
      & 1999 & 0.056 $\pm$ 0.004 & 21.487 &   13.15 &  2.38 & 1.00 & -105.8  & 1.28 \\
      & 2003 & 0.068 $\pm$ 0.004 & 21.613 &   13.42 &  2.38 & 1.00 & -105.8  & 2.07 \\
\enddata
\tablenotetext{*}{NOTE - Parameters of each Gaussian component of the model
brightness distribution are as follows:  Component, Gaussian component (see Figure \ref{VLAmaps}); 
Epoch, year of observation (see Table \ref{Observations} and \S2);  
S, flux density; r, $\theta$, polar coordinates of the
center of the component relative to the center of component C1; a, semimajor axis; b/a, axial ratio; 
$\Phi$, component orientation; $\chi^{2}$, goodness-of-fit for eight component model in each epoch. 
Polar angles are measured
from north through east.  Errors in flux are based on our absolute amplitude calibration
as well as the rms noise.  Note that due to the complicated morphology of the source, variability
studies were performed using components C1, C2, the sum of the southern components (S1, S2,
S3, and S4), and the sum of the northern components (N1 and N2).}
\end{deluxetable}

\begin{deluxetable}{ccccc}
\tabletypesize{\scriptsize}
\tablecolumns{6}
\tablewidth{0pt}
\tablecaption{Component Motion Fitting Results.\label{Motions_Table}}
\tablehead{\colhead{Component}&\colhead{Velocity}&\colhead{Projection Angle\tablenotemark{*}}
&\colhead{$\chi^{2}$}&\colhead{Status} \\
\colhead{} & \colhead{(mas/yr)} & \colhead{(degrees)} & \colhead{} & \colhead{}}
\startdata
C1 & ... & ... & ... & Reference\tablenotemark{+} \\
C2 & 0.0468 $\pm$ 0.019 &  241.48 & 4.05  & 0.161 $\pm$ 0.065 \\
S1 & 0.0727 $\pm$ 0.010 & -143.31 & 33.95 & No motion \\
S2 & 0.0034 $\pm$ 0.010 &  -93.72 & 1.28  & No motion \\
S3 & 0.0124 $\pm$ 0.010 &  197.42 & 0.08  & No motion \\
S4 & 0.0047 $\pm$ 0.010 &  190.58 & 4.26  & No motion \\
N1 & 0.1670 $\pm$ 0.010 &   14.58 & 72.21 & No motion \\
N2 & 0.0398 $\pm$ 0.010 &   54.82 & 1.38  & 0.137 $\pm$ 0.034 \\
\enddata
\tablenotetext{*}{Angles measured from north through east.}
\tablenotetext{+}{Position $\sim$ RA 04h05m49.2623s Dec +38d03m32.235s (Beasley et al. 2002)}
\end{deluxetable}

\begin{deluxetable}{lcccccc}
\tabletypesize{\scriptsize}
\tablecolumns{7}
\tablewidth{0pt}
\tablecaption{Continuum Spectrum Results\tablenotemark{*}.\label{Peak_Fluxes}}
\tablehead{\colhead{Component}&\colhead{Frequency}&\colhead{Epoch}
&\colhead{Peak Flux}&\colhead{RMS Noise}&\colhead{$\alpha_{1-5}$}
&\colhead{$\alpha_{5-15}$}\\
\colhead{} &\colhead{(GHz)}&\colhead{} &\colhead{(mJy/beam)} 
&\colhead{(mJy/beam)}&\colhead{}&\colhead{}}
\startdata
C1... &  1.347 & 2003 & unresolved &     & ...   & 0.20  \\
      &  2.220 & 1996 & 26.38      & .94 &       & \\
      &  4.991 & 2003 & 53.53      & 2.1 &       & \\
      &  8.150 & 1996 & 26.92      & 2.4 &       & \\
      &  15.36 & 2003 & 66.72      & 1.3 &       & \\
C2... &  1.347 & 2003 & unresolved &     & ...   & 0.35 \\   
      &  2.220 & 1996 & unresolved &     &       & \\
      &  4.991 & 2003 & 12.79      & 2.1 &       & \\
      &  8.150 & 1996 & 10.12      & 2.4 &       & \\
      &  15.36 & 2003 & 18.89      & 1.3 &       & \\
S2... &  1.347 & 2003 & 247.89     & .49 & -0.42 & -1.38 \\
      &  2.220 & 1996 & 351.65     & .94 &       & \\
      &  4.991 & 2003 & 141.92     & 2.1 &       & \\
      &  8.150 & 1996 & 76.33      & 2.4 &       & \\
      &  15.36 & 2003 & 29.93      & 1.3 &       & \\
N2... &  1.347 & 2003 & 162.97     & .49 & -1.32 & -1.60 \\       
      &  2.220 & 1996 & 104.48     & .94 &       & \\
      &  4.991 & 2003 & 28.68      & 2.1 &       & \\
      &  8.150 & 1996 & 7.91       & 2.4 &       & \\ 
      &  15.36 & 2003 & 4.74       & 1.3 &       & \\
\enddata
\tablenotetext{*}{See \S 3.3 for table comments.}
\end{deluxetable}

\begin{deluxetable}{ccccc}
\tabletypesize{\scriptsize}
\tablecolumns{6}
\tablewidth{0pt}
\tablecaption{Gaussian Function Fitted To H{\scriptsize I} Absorption Profiles In Region S1.\label{HItab}}
\tablehead{\colhead{Amplitude}&\colhead{Central Velocity}&\colhead{FWHM}
&\colhead{$\tau$}&\colhead{N$_{H{\scriptsize I}}$\tablenotemark{*}}\\
\colhead{(mJy)} &\colhead{(km s$^{-1}$)} &\colhead{(km s$^{-1}$)} 
&\colhead{} &\colhead{($\times$ 10$^{20}$ cm$^{-2}$)}}
\startdata
5.2 $\pm$ 0.49 & 17,115 $\pm$ 23 & 540 $\pm$ 69 & 0.0179 $\pm$ 0.0007 & 1.8 $\pm$ 0.2 \\
\enddata
\tablenotetext{*}{Assuming a spin temperature of 100 K and a covering factor of 1.}
\end{deluxetable}
\clearpage 

\end{document}